\documentstyle[12pt,a4,graphicx]{article}

\newcommand{\be}{\begin{equation}}
\newcommand{\ee}{\end{equation}}
\newcommand{\ba}{\begin{eqnarray}}
\newcommand{\ea}{\end{eqnarray}}

\newcommand{\re}{\mbox{Re}\,}
\newcommand{\im}{\mbox{Im}\,}

\begin{document}
\begin{titlepage}
\begin{flushright}
CAFPE-39/04\\
UB-ECM-PF-04-27\\
UG-FT-169/04\\
\end{flushright}
\vspace{2cm}
\begin{center}

{\large\bf Charged Kaon $K\to3\pi$ CP Violating Asymmetries
\footnote{
Invited talk given by J.P.
at `` 32nd ICHEP, 16-22 August 2004, Beijing, China''.}}\\
\vfill
{\bf Elvira G\'amiz$^{a)}$, 
 Joaquim Prades$^{b)}$ 
and  Ignazio Scimemi$^{c)}$ 
}\\[0.5cm]
$^{a)}$ Department of Physics \& Astronomy,
University of Glasgow\\ Glasgow G12 8QQ, United Kingdom.\\[0.5cm]

$^{b)}$ Centro Andaluz de F\'{\i}sica de las Part\'{\i}culas
Elementales (CAFPE) and Departamento de
 F\'{\i}sica Te\'orica y del Cosmos, Universidad de Granada \\
Campus de Fuente Nueva, E-18002 Granada, Spain.\\[0.5cm]

$^{c)}$ Facultat de F\'{\i}sica-ECM, Universitat de Barcelona\\
Av. Diagonal 647, E-08028 Barcelona, Spain.\\[0.5cm]
\end{center}
\vfill
\begin{abstract}
\noindent
First full next-to-leading order analytical results
in Chiral Perturbation Theory for the charged Kaon
$K\to 3\pi$ slope $g$ and decay rates CP-violating
asymmetries are presented. We discuss the constraints
that a  measurement of these asymmetries would impose
on the Standard Model calculations of $\varepsilon_K'$
and the kind of information it can provide on 
Im $G_8$, Im $(e^2 G_E)$ and higher order weak couplings.

\end{abstract}
\vfill
October 2004
\end{titlepage}

\section{Introduction and Motivation}

Direct CP violation has been established unambiguously in
$K\to \pi\pi$ decays by KTeV\cite{KTeV} and NA48\cite{NA48}
through the measurement of Re($\varepsilon_K'/\varepsilon_K$).
Its present world average is\cite{KTeV,NA48,NA31,E731}
\be
\label{epsilon}
{\rm Re} \left( \frac{\varepsilon_K'}{\varepsilon_K} \right)
=(1.67\pm0.16) \cdot 10^{-4} \, .
\ee

The theoretical understanding of this quantity within 
the Standard Model (SM) is not at the same level.
We mention here just the most recent advances:
 the Chiral Perturbation
Theory (CHPT) calculation\cite{KMW90,K2piNLO}   and the 
isospin breaking corrections\cite{CENP03} have both fully been done at 
next-to-leading order (NLO) and  the r\^ole of Final State Interactions
(FSI) has also been understood\cite{FSI}
--for a more extensive description
of these works  and references, see\cite{Toni}.
 There have been also recent 
advances on the calculation of the leading-order (LO) CHPT 
couplings Im $G_8$ and Im ($e^2 G_E$)
\cite{elmatrix,strmatrix,matrix,latmatrix}
--they are not fully under control though and
 more work is still needed.

Asymmetries in the Dalitz variable slope $g$
of $K\to 3\pi$ amplitudes are  another very promising place
to study direct CP violation in Kaon decays. In fact, there are several
experiments, NA48/2\cite{KEK04} at CERN, KLOE\cite{KLOE}
 at Frascati and OKA\cite{OKA}
at Protvino, that have announced an expected sensitivity to these
asymmetries
of the order of $10^{-4}$, one order of magnitude better than at 
present\cite{AJI03}.  On the theory side,  though the first 
calculation of $K \to 3\pi$ at NLO  in CHPT was done long ago\cite{KMW90},
the analytical full results  were unfortunately not available
until recently\cite{BDP03}. The CP asymmetries were therefore 
predicted just at LO plus various
estimates of NLO effects\cite{previous}. 
The first full NLO calculation within CHPT for those  asymmetries
was done in \cite{GPS03}. Here, we report 
results just for $\Delta g_C$ --results
for the rest of the asymmetries can be found.

\section{Technique}

The effective quantum field theory of the SM at energies
below or of the order of 1 GeV is CHPT\cite{CHPT}.
Some introductory lectures on CHPT can be found in
\cite{lectures} and recent reviews in\cite{reviews}.

 The full one-loop calculation in the isospin
limit was done in \cite{BDP03,GPS03} and they both  fully agree.
 All the needed  notation and definitions were given there.
 Recently, some isospin breaking corrections have also been 
calculated\cite{isoK3pi}.
Notice that  some  misprints in the first reference in\cite{GPS03} 
were reported in the third reference in\cite{GPS03}. 

At this order there appear eleven unknown counterterms.
The real part of them and of the LO  couplings $G_8$ and $G_{27}$
can be fixed from a fit to all available $K\to \pi\pi$ amplitudes
at NLO in CHPT\cite{K2piNLO} and $K\to 3\pi$ amplitudes and slopes
also at NLO\cite{BDP03,GPS03}. This was done in \cite{BDP03}
 and we used them  as inputs in all the results we report here.

The values we used for  Im $(e^2 G_E)$  and Im $G_8$ 
can  be  found in\cite{GPS03}. They are  taken  mainly
from \cite{elmatrix,strmatrix}
but are also compatible with \cite{matrix,latmatrix}.

The imaginary part of the order $p^4$ counterterms, Im $\widetilde K_i$,
 is much more problematic. They cannot be obtained from data and
there is no available calculation for them at NLO in $1/N_c$.
One   can still get the order of magnitude and/or signs of
Im $\widetilde K_i$ using several approaches. We  followed\cite{GPS03}
a more  naive approach that is enough for the purpose of estimating the
effect of those counterterms. We assumed that the ratio of the
real to the imaginary part is dominated by the same strong dynamics
at LO and NLO in CHPT, namely
\ba
\frac{\im \widetilde K_i}{\re \widetilde K_i}
&\simeq& \frac{\im G_8}{\re G_8} 
\simeq \frac{\im G_8'}{ \re G_8'} \simeq (0.9 \pm 0.3) \, \im \tau
\nonumber \\ 
&=& - (0.9\pm0.3) \,\,
\im \left(\frac{V_{td}V_{ts}^*}{V_{ud}V_{us}^*}\right).
\ea

\section{$K\to 3\pi$ CP Violating Asymmetries}
 
The definition of the CP-violating asymmetries in the slope $g$ 
and analogous asymmetries for the decay rates $\Gamma$
can be found, for instance, in\cite{GPS03}.
They  start at ${\cal O}(p^2)$ in CHPT and at NLO require
the FSI phases of three-pions at NLO, i.e. an 
${\cal O}(p^6)$ calculation.

Though the full result is unavailable at present, 
we have calculated
analytically the expected dominant part
which comes from  two-bubble diagrams\cite{GPS03}. Including these
and substituting the pion and Kaon masses,
Re $G_8$, $G_{27}$  and the real part of the NLO 
CHPT couplings, the result we get for $\Delta g_C$ is 
\ba
\label{deltagc}
\frac{\Delta g_C}{10^{-2}} &\simeq&
\left[ \left(0.7\pm0.1\right) {\rm Im} G_8 + \left(4.3\pm 1.6\right) 
 {\rm Im} \widetilde K_2 \right.  \\
 &-& \left.(18.1 \pm 2.2) {\rm Im} \widetilde K_3
-(0.07 \pm0.02) {\rm Im} (e^2 G_E) \right] \, . \nonumber
\ea
When values for the imaginary part of the needed couplings 
are taken as explained in the previous section, on gets
\be
\Delta g_C=-(2.4\pm1.2) \cdot 10^{-5} \, .
\ee
Results for the rest of the asymmetries can be found
in\cite{GPS03}

\section{$\varepsilon_K'$ vs $K\to3\pi$ CP Violating Asymmetries}

\begin{figure}
\begin{center}
\begin{minipage}[t]{1cm}\vskip-4.cm 
$\frac{{\rm Im} G_8}{{\rm Im} \tau}$\end{minipage}
\hspace*{-.5cm}\includegraphics[height=0.45\textwidth]{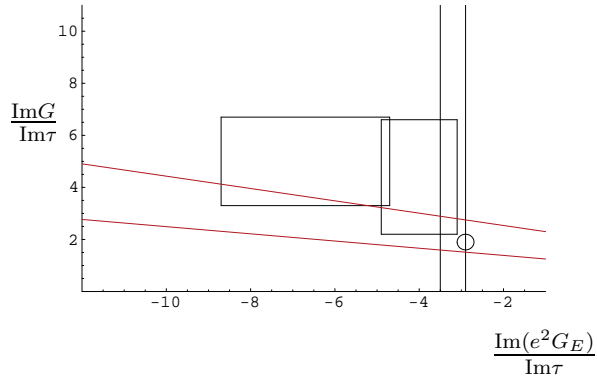}
\begin{minipage}[b]{1cm}\hspace*{-1cm}\vspace{.4cm}
$\frac{{\rm Im} (e^2 G_E)}{{\rm Im} \tau}$\end{minipage}
\caption{$\varepsilon_K'$: Theory vs Experiment. See text for explanation.}
\label{fig:status}
\end{center}
\end{figure}

Including  FSI to all orders, 
CHPT and isospin breaking at NLO\cite{K2piNLO,CENP03,FSI},
one gets
\be
\frac{\varepsilon_K'}{\varepsilon_K} \simeq
-\left[ \left( 1.88\pm1.0\right) {\rm Im} G_8 +  \left(0.38\pm0.13
\right) {\rm Im} (e^2 G_E) \right] \, .
\ee
Using this result, the experimental one in (\ref{epsilon}) imposes
that Im $G_8$ and Im $(e^2 G_E)$ are constrained
to be within the  horizontal band in Figure \ref{fig:status}.
Also plotted in the same figure are the predictions for those
couplings from \cite{elmatrix,strmatrix} --rectangle on the right--, 
 from \cite{matrix} --rectangle on the left--
and from \cite{latmatrix} --vertical lines.

\begin{figure}
\begin{center}
\begin{minipage}[t]{1cm}\vskip-4.cm 
$\frac{{\rm Im} G_8}{{\rm Im} \tau}$\end{minipage}
\hspace*{-.5cm}\includegraphics[height=0.45\textwidth]{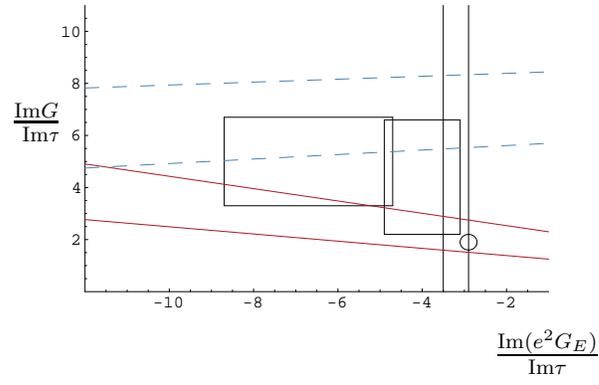}
\begin{minipage}[b]{1cm}\hspace*{-1cm}\vspace{.4cm}
$\frac{{\rm Im} (e^2 G_E)}{{\rm Im} \tau}$\end{minipage}
\caption{$\varepsilon_K'$ vs $\Delta g_C$ for  $\Delta g_C =
- 3.5 \cdot 10^{-5}$.}
\label{fig:gclim}
\end{center}
\end{figure}

\begin{figure}
\begin{center}
\begin{minipage}[t]{1cm}\vskip-4.cm 
$\frac{{\rm Im} G_8}{{\rm Im} \tau}$\end{minipage}
\hspace*{-.5cm}\includegraphics[height=0.45\textwidth]{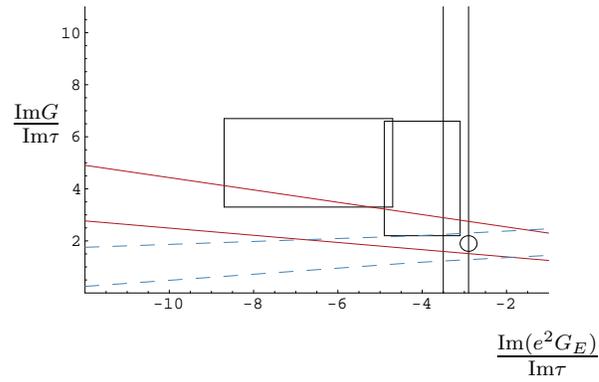}
\begin{minipage}[b]{1cm}\hspace*{-1cm}\vspace{.4cm}
$\frac{{\rm Im} (e^2 G_E)}{{\rm Im} \tau}$\end{minipage}
\caption{$\varepsilon_K'$ vs $\Delta g_C$ for  $\Delta g_C =
- 1 \cdot 10^{-5}$.}
\label{fig:figc}
\end{center}
\end{figure}

A measurement of $\Delta g_C$
can have an important impact on constraining what we know on
Im $G_8$ and and Im $(e^2 G_E)$ from $\varepsilon'_K$.
To assess the quality of these constraints, we plot
in Figure \ref{fig:gclim}  the comparison between what one gets
with $\varepsilon_K'$, the theory predictions
and the dashed horizontal band that
one gets using (\ref{deltagc})  for  $\Delta g_C = -3.5 \cdot 10^{-5}$.
In Figure \ref{fig:figc}, we show the same plots for    
$\Delta g_C = -1 \cdot 10^{-5}$.

\section{Conclusions}
 
 The CP violating asymmetry $\Delta g_C$ is dominated by the value of
Im $G_8$  and its final uncertainty is mainly from this input. 
This is the only
asymmetry with an uncertainty smaller than 50\%. The predictions for
the rest of CP asymmetries can be found in \cite{GPS03}.

The eventual measurement of  $\Delta g_C$ will then provide a check
of consistency with $\varepsilon_K'$ --see Figures 
\ref{fig:gclim} and \ref{fig:figc}.
The SM prefers values  for
 this asymmetry larger than $-0.4\cdot 10^{-4}$ and 
an experimental result of the order or smaller than $-2 \cdot 
10^{-4}$ would indicate the presence of new physics.
For a discussion on possible SUSY implications of a measurement
of these asymmetries see \cite{AIM00}.

 The CP asymmetries $\Delta g_N$ and in the decay rates were
also discussed in \cite{GPS03} and we found that they are dominated by
the imaginary part of the ${\cal O} (p^4)$ counterterms.
 A measurement of these asymmetries would therefore give
very interesting information on the size of the imaginary parts
of those couplings.

As a general conclusion,
direct CP violating asymmetries  in $K\to 3 \pi$
 provide extremely  interesting and valuable
information on the SM which is complementary 
to the one obtained from $\varepsilon_K'$.
We are therefore eagerly awaiting the new experimental results!

\section*{Acknowledgments}
E.G. is indebted to the EU for a Marie Curie Fellowship.
This work has been supported in part by the EU
RTN Network EURIDICE under Contract No. HPRN-CT2002-00311 
(I.S. and J.P.), by MEC (Spain) and FEDER (EU)
Grants No. FPA2001-03598 (I.S.)  and  FPA2003-09298-C02-01 (J.P.)
and by Junta de Andaluc\'{\i}a Grant No. FQM-101 (J.P.).

\end{document}